\documentclass[12pt]{iopart}

\usepackage{graphicx} 
\usepackage{bbm}
\usepackage{amssymb}

\begin{document}

\title[The emergence of bluff in poker-like games]{The emergence of bluff in poker-like games}

\author{Andrea Guazzini$^1$ and Daniele Vilone$^{2}$}
\address{$^1$ Institute for Informatics and Telematics, National Research Council (CNR), Pisa, Italy}
\address{$^2$ Center for the Study of Complex Dynamics, Universit\`a di Firenze, Florence, Italy}
\ead{andrea.guazzini@gmail.com, daniele.vilone@gmail.com}

\begin{abstract}
We present a couple of adaptive learning models of poker-like games, by means of which we show
how bluffing strategies emerge very naturally, and can also be rational and evolutively stable.
Despite their very simple learning algorithms, agents learn to bluff, and the best
bluffing player is usually the winner.
\end{abstract}

PACS: 02.50.Le, 89.65.-s

\section{General introduction}

Among the concepts formulated and developed by psychology together with neurosciences in the last century, that of cognitive process has enabled a significant improvement in the comprehension of the mental activity and of its relations with cerebral circuits. 

Initially  the cognitive process was interpreted solely in terms of emerging object at a certain level of observation of the mental activity, and therefore equipped with an axiomatic characterization mostly anchored to the properties empirically observed though psychological searches.

At this descriptive level the cognitive process represented the necessary link at theoretical level  between the superior cognitive functions and the human mental activity. As a matter of fact the idiosyncrasy was emerging and certain abilities of the cognitive system such as language, perception and memory, were taken into consideration as key components of the cognitive strategies used by human beings in everyday life. 

In short these last ones did not show a functional stability similar to that of canonical cognitive functions, they did not show critical periods in the ontological development during which an incredible plasticity and sensitiveness to the environmental determinants emerged and seemed to be in continuous development. 

Concluding even if the concept of cognitive function turned out to be precious in order to explain the emerging of the basic and innate abilities of the human brain of computing the environmental information, these did not suffice to represent neither the complexity of the strategies employed by human beings to solve complex problems, nor how such functional structures could develop in the space of their life towards stable and efficient strategies called \textit{problem soliving strategies}.

On the wave of such necessity and also tanks to the then recent discoveries of neurosciences about the organization of the neuro-cognitive system, psychology stared to define through an empiric-inductive method the characteristics of a superior organization of mental activity based on cognitive functions. 

For such  purpose the concept of cognitive process wad defined both in terms of result of the parallel elaboration of several well defined and functionally independent neural-moduli, and in terms of a software able to optimize the integration between the activity of different cognitive functions by adapting to different environmental/informational circumstances. 

The concept of cognitive process simultaneously solved two theoretical dead ends which afflicted neurosciences since their birth. 

The first one concerned the dichotomy: physiological learning-development. In fact while the hebbian principle properly explained how the neural activity could evolve discrete spatial structures as the product of the neural activity itself, thus laying the basis for a comprehension of the development of the human innate  neural circuits, this did not suffice anyway to clarify the mechanism which lay  at the basis of the development of the general rules which characterized the processes of learning itself, though it was a fundamental ingredient of it. 

The second impasse divided the studies on human 'creativity' and in general on the complex analysis of environmental information which seemed to characterize mental activity, from the more solid studies on the neural organization and functioning. The physiological basis of knowledge still  represented an 'utopia' and the extraordinary ability of man to adapt to different situation through the structuring in a seemingly creative way of clever solutions to current problems represented still a mystery. 
 
The study of human behaviour in properly standardized condition, yet without relegating it to the principal setting of perceptological tradition, allowed to grasp that the dyad fundamental to understand and reduce the variety of mental production was the one constituted by environment and individual.
Considering the environment as an object of search it was actually possible to individuate several regularities which characterized the strategies of analysis of information used by individuals in different situations. Among the various disciplines Social cognition is perhaps the one that has formalized the most interesting taxonomy of cognitive processes though the use of this paradigm.  Concepts such as heuristics, metal scheme, problem solving, decision making, mental inference and so on, seem not only to be building blocks fundamental to explain  the superior mental activity, but they also represent today concepts which make sense in  the search of neurosciences.  

Finally  the recent coming back of such theoretical representations and the massive attention paid to them by very different scientific domains, is for the most part due to the possibility of describing  in computational terms the structure of such processes.
This kind of searches has stressed how some algorithms optimized to reduce the entropy of environmental information, characterized by an extraordinary  adaptation skill and able to develop even very different functional structures, result from the cognitive strategies of information analysis.

\section{Poker-like games}

Among others poker is an interesting test-bed for artificial intelligence research \cite{Davidson2002,Koller1997,Johansen2001,Barany1983}. It is a game of imperfect information, where multiple competing agents must deal with probabilistic knowledge, risk assessment, and possible deception, not unlike decisions made in the real world. Opponent modeling is another difficult problem in decision-making applications, and it is essential to achieving high performance in poker.
Moreover poker has a rich history of study in other academic fields. Economists and mathematicians have applied a variety of analytical techniques to poker-related problems \cite{Zhizhang2005,Rapoport1997,Miekisz2007}. For example, the earliest investigations in game theory, by luminaries such as John von Neumann and John Nash, used simplified poker to illustrate the fundamental principles.

However, there is an important difference between board games and popular card games like bridge and poker. In board games, players have complete knowledge of the entire game state, since everything is visible to both participants. In contrast, bridge and poker involve imperfect information, since the other players' cards are not known. 

\ 

\begin{center}
\begin{tabular}{|c|c|c|}
\hline
	&	Perfect information	&	Imperfect     information \\
\hline
No chance  &     Chess            &          Inspection game \\
\hline
Chance	 &      Monopoly     &                     Poker \\
\hline
\end{tabular}\end{center}

\ 

From a computational point of view it is important to distinguish the lack of information from the possibility of chance moves. The former involves uncertainty about the current state of the world, particularly situations where different players have access to different information. The latter involves only uncertainty about the future, uncertainty which is resolved as soon as the future materializes. Both perfect and imperfect information games may involve an element of chance; examples of games from all four categories are shown in Figure. 

The presence of chance elements does not necessitate major changes to the computational techniques used to solve a game. In fact, the cost of solving a perfect information game with chance moves is not substantially greater than solving a game with no chance moves. By contrast, the introduction of imperfect information greatly increases the complexity of the problem. 

Due to the complexity (both conceptual and algorithmic) of dealing with imperfect information games, this problem has been largely ignored at the computational level untill the introduction of randomized strategies concept.

Once randomized strategies are allowed, the existence of 'optimal strategies' in imperfect information games can be proved. In particular, this means that there exists an optimal randomized strategy for poker, in much the same way as there exists an optimal deterministic strategy for chess. Indeed, Kuhn has shown for a simplified poker game that the optimal strategy does, indeed, use randomization \cite{Kuhn1950}.

The optimal strategy has several advantages: the player cannot do better than this strategy if playing against a good opponent; furthermore, the player does not do worse even if his strategy is revealed to his opponent, i.e., the opponent gains no advantage from figuring out the player's strategy. 

Another interesting results of this researches face with the optimal strategies for the gambler in poker game. As first observed in a simple poker game by Kuhn, behaviors such as bluffing, that seem to arise from the psychological makeup of human players, are actually game-theoretically optimal.

One of the earliest and most thorough investigations of Poker appears in the classical treatise on game theory: Games and Economic Behavior by von Neumann and Morgenstern \cite{VonNeumann1947}, where a large section was devoted to the formal analysis of 'bluffing' in several simplified variants of a two-person Poker game with either symmetric or asymmetric information. 

Indeed, the general considerations concerning Poker and the mathematical discussions of the variants of the game were carried out by von Neumann as early as 1926. Recognizing that 'bluffing' in Poker 'is unquestionably practiced by all experienced players,' von Neumann and Morgenstern identified two motives for bluffing. 'The first is the desire to give a (false) impression of strength in (real) weakness; the second is the desire to give a (false) impression of weak-ness in (real) strength' \cite{VonNeumann1947}.

Solutions to these variants of the simplified Poker game as well as a large class of both zerosum and nonzerosum games were unified by the concept of mixed strategy, a probability distribution over the player's set of actions. The importance of mixed strategies to the theory of games and its applications to the social and behavioral sciences stems from the fact that for many interactive decision situations modeled as noncooperative games, both zerosum and nonzerosum, there are no
Nash equilibria in pure strategies.

Using randomization and adaptive learning as key concepts to modelize into a computational scaffholding the cognitive processes, we believe that this area of research is more likely to produce insights about superior cognitive strategies because their intrinsically structures which allow to represents effectively the interactions between task's informational contents and evolution of optimal stable strategies of information analysis.
Finally comparing them with the real human's strategy it is possible both to investigate the role of environmental factors on cognitive strategies development and to validate theoretical psychological assumptions.

\ 

\subsection{Summary}

In conclusion the targets of this work may be expressed through three different challenges:

The first target is: to individuate the fundamental set of rules that starting from the known and measurable cognitive functions allow the development of rational strategies for the poker. In other words: is it possible to built a computational system anchored to the constraints defined by neurosciences and psychology able to apprehend strategies near to those actually observed in poker players?

The second is: may counterintuitive behaviours such as bluff emerge in such systems in terms of rational and evolutionally stable strategies? 

And eventually, does the speed at which the system adapts, and therefore develops new strategies the basis of experience, represent a key parameter for the fitness of the individual in such circumstances?

The work has been divided in two fundamental parts, one is  the qualitative inspection of the model, and the other one is the mathematical analysis. 

In the first part a cognitive sciences-inspired formal representation of the surveyed model is proposed , by studying its macroscopic phenomenology through numerical simulations with agent based system using a simplification of the Italian poker. 
The model represents each agent through three probability matrixes associated to the respective possible actions (e.g. Raise, Call and Fold) and through the evolution rules of such probability matrixes.
Finally each agent was also characterized  by another parameter, called Learning Factor LF, which represents the speed at which each agent modifies its strategy during the game. Such parameter allows to highlight both how much the dynamic characteristics of the learning process weigh on the fitness of the subject, and in short how these represent the fundamental emerging characteristic of the cognitive processes.

\ 

In the second part of the chapter, we present an even more simple model of poker-like game, with only two players,
and only two possible strategies: folding and calling, which each agent assumes simultaneously.
Such oversimplified game, as we will see, allows to catch the fundamental mechanisms underlying the phenomenon of
bluffing. The fact that bluffing naturally emerges already in a very simple version of the game seems to suggest
that such a strategy is perfectly rational, and can be mathematically characterized.

\section{First part: the four agent model}

\subsection{Implementation of a simplified version of poker card game}

Poker is a large family of games with similar rules. For any given form of poker there is a set of technical syntactic rules, which control the flow of the game. These rules specify the number of cards that should be dealt, the order in which the players make their bets, and so on. The syntactic rules fit the game-theoretic framework perfectly, and one can easily produce an algorithm that implements them for any given form of poker.

The first step of the survey was therefore that of formalizing the form of poker which could be implementable in a multi agent based system. It was decided to simplify some of its characteristics, yet without altering in any major way the structure of the game.

Finally the simplified poker game here proposed is a four-agents game and it can be described in terms of six steps as follows:

\begin{itemize}
 \item Step 1. Each of four players puts a single poker chip in the pot.
\item Step 2. A random number between 0 and 1 is generated in order to represents the absolute value of the 'hand' for each player. No change of cards is allowed and this value represents the final score of each player.   
\item Step 3. The first player, which is different for each hands, makes one of two decisions called 'Call' (C) and 'Raise' (R). Only for the first player of each hand folding is not allowed. If player 1 calls he put no money into the pot and next player turn begin. While if he raises, player 1 adds one more chips to the pot and pass the turn. 
\item Step 4. If no players raise the show down phase begins. If one player raises next players can 'Fold', 'Call' or 'Raise". If all fold, raiser player gets money inside the pot and game ends. If all call everyone put a chip into the pot and show down phase begins. Otherwise one player can raises again putting 2 chips on the pot, in this case step 4 is repeated.
\item Step 5. In the show down phase the player which has the grater hand value gets money.
\item Step 6. In final step player strategy is updated according with model rules.
\end{itemize}

\subsection{The model: a cognitive based formalization of poker player's strategies and adaptive learning rules.}

The following step was that of associating to each agent (i.e. virtual player) a game strategy and several rules of evolution which could allow him not only to learn how to better play but also to take into consideration the environmental variables represented by others players and their strategies.
Resorting to statistical Mechanics  a normalized probability was associated to each one of the three possible actions (i.e. calling folding and raising) bearing in mind both the hand value and the pot size. It was therefore necessary to define three distinct matrixes for each agent, normalizing for all the three of them the values of the cells corresponding to $1$.

Finally through a montecarlo method such matrixes have properly represented the game strategy of the agents. 
Yet the key ingredient of the model at this level do not lay much in the formal representation of the problem solving process as it does in its rule of evolution. In fact as reminded at the beginning the peculiarity of cognitive processes is not only that of structuring reliable algorithms for a given circumstance, but it lays mostly in their ability to change and evolve different structures depending on the environmental conditions.  

In this sense the absence on the market of models which explicitly characterize in cognitive terms such aspect has pushed us to implement a very simple set of rules inspired to the so-called cognitive heuristics. In particular we have had recourse to the concept of representativeness heuristics in order to justify the existence of an anchoring of the probability of  a determined action to the previous experience (the probability matrixes).  The availability heuristics~\cite{guazth} instead was considered in order to represent the dynamic of the learning process. In this sense the probability associated to the last made action is either updated and incremented when it is a winning one or reduced if it a losing one, furthermore the shift of strategy is modulated by the entity of the achieved loose/win in relation to the previous capital. The sole exception is represented by the fold strategy that is each time reformulated though a renormalization with respect to the other two, because in this case the system would anyway evidently tend to diminish the probability of the fold since this last causes a constant and inevitable loss. 

\subsubsection{A simple evolutionary rules}

\begin{equation}
\label{}
P(h,p,d)^t_{i}=P(h,p,d)^{t-1}_{i}+\mu_{i}\cdot\frac{((Chips_{i}^t-Chips_{i}^{t-1})}{Chips_{i}^{t-1}}\\
\end{equation}

$\forall$ given hand value ($h$) and pot size ($p$) $\rightarrow P(Fold)+P(Call)+P(Raise)=1$

\ 

Thus each agent is characterized by an added parameter , labelled learning factor, comprised between 0 and 1. In effect such parameter determines both the learning speed of the agent and its inclination towards risk, thus regulating the order of greatness of the probability variation determined by the learning process. 

\subsection{Numerical Simulations}
The simple model described above has been implemented in matlab and studied by numerical simulation in order to investigate the following aspects:

\begin{itemize}
 \item Are these rules sufficient to evolve a rational and effective strategy for poker game?
\item Is the bluff an emergent phenomenon reproduced by this approach? 
\item What is the role of the learning factor?
\end{itemize}

To answer to these questions several campaigns of simulations, each of them constituted by 20 games of 2000 hands has been perfomed. Moreover both systems of agents characterized by the same Lfs and systems where an agent had a different LF from others has been considered. Finally different values of LF for both these conditions has been also investigated.
At the begininning of each game 1000 poker chips were assigned to each player and each matrices elements were inizialized at a constant value of 1/3. At the end of each game the amount of money for each players and the developed strategy were registered, and the initial conditions restored (REF appendice).
Finally in order to describe the system the mean of stable strategy developed in different games and the trend of agent's money (chips) have been assumed as order parameter, while LF has played the role of control parameter. 

\subsubsection{Simulation results}

Results section is organized in two distinct part for: the first one considers results of the LF-homogeneous system, investigating both final players' strategies characteristics, and the trend of money for each player for different values of LF. Second section presents the same results for the LF-heterogenous system, and finally focuses on the role of LF on the agent's fitness.

\subsubsection{LF-Homogeneous system}

\begin{figure}
\centering
\includegraphics[width=12cm,height=6cm]{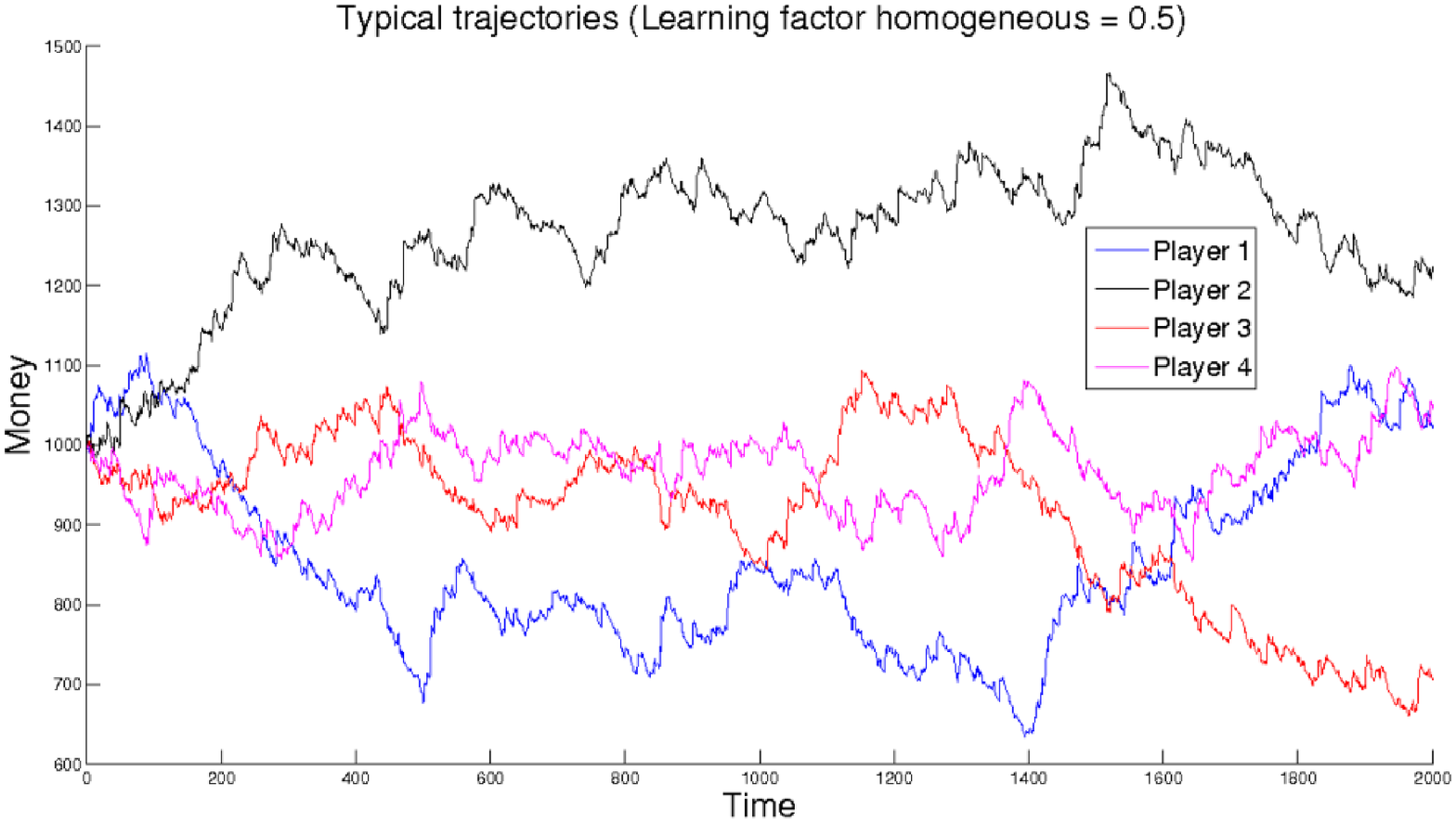}
\caption{}\label{fig:typtraj1}
\end{figure}

In this first condition all agents were characterized by the same LF,  fixed in a consevative way at 0.5. At the beginning of each game all elements of strategie's matrices were fixed at 1/3 in order to eliminitate any difference in among agents in initial conditions.
A typical trajectory of a single game for this condition is showed in \ref{fig:typtraj1}

\begin{figure}
\centering
\includegraphics[width=15cm,height=13cm]{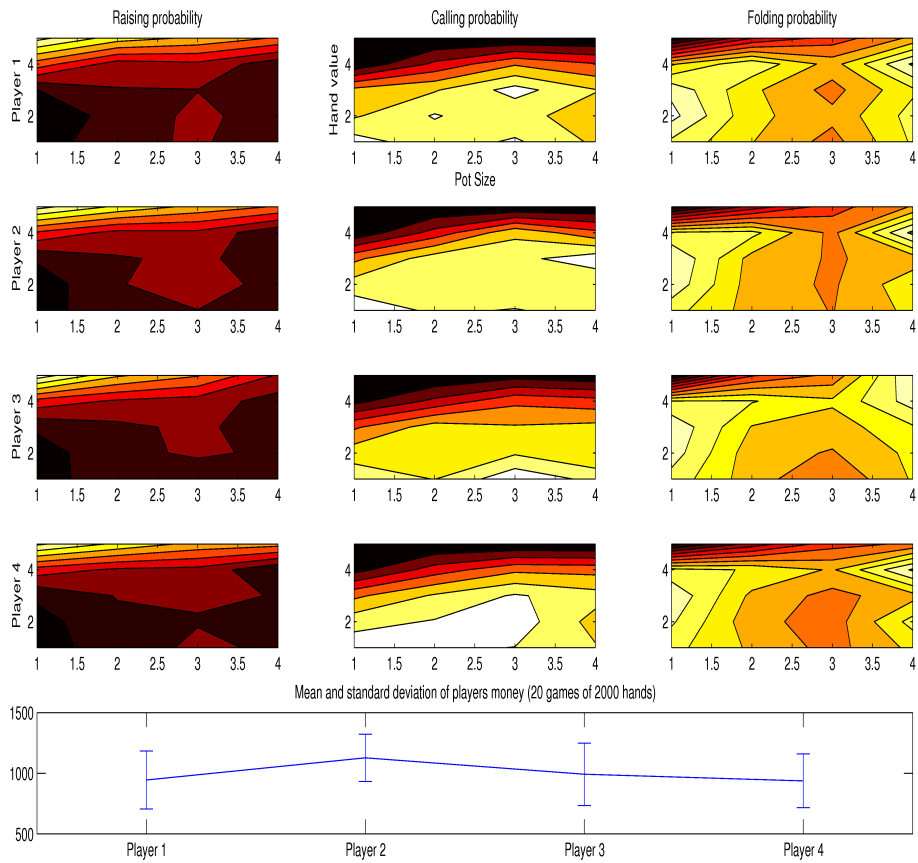}
\caption{For each player the three matrices which define the player's stable strategy are obtained considering the mean among 20 different games of 2000 hands. In the bottom of the figure a mean and standard deviation of player's money in the final state is also presented}\label{fig:Multy05}
\end{figure}

The figure \ref{fig:typtraj1} shows the progress over time of the chips for each agent. As it will be subsequently evident the relatively restricted range of variation of the players' chips, comprised in its maximum  size between around $700$ and $1400$, shows how none of them dominates the others.
In order to confirm such hypothesis it is necessary to analyze the average gain of each player on several different simulations, each one of which produced starting from the same initial conditions. 
Ad shown in the lower part of figure \ref{fig:Multy05} the players' average gain in this experimental condition is near to $0$ in the space of $20$ games and the standard deviation shows similar values for the four players and of the order of a quarter of the initial capital.

Finally the analysis of the average of the probability matrix associated to the strategies developed by the players during the $20$ games of the simulation is of particular interest. At a first analysis it is possible to notice how all the players develop similar conformations of the three matrixes.
Taking  into consideration the raising probability matrix of the player number $2$ it is possible to notice as well interesting and particularly evident characteristics. In this the probabilities near to $1$ are indicated in light colours and in dark colours the ones near to $0$. It is immediately possible to notice that this agent tends to raise more frequently with an high score than it does with a lower one, and basically more frequently with small pots than with big ones, exactly as we would expect a real player to do.

Furthermore a more accurate analysis shows some characteristics, noticeable as well in the other players, which make us think of the \textit{bluff}.  In effect we would expect that with an intermediate score on hand the probabilities of raising are a monotonous decreasing function of the pot size, while  discontinuities of this kind are noticeable in all the three matrixes of each player. Different campaigns of simulation have been repeated in order to verify that such characteristics were not due to particular regimes, but extraordinary coincidences have always characterized the spaces of probability thus obtained.

\begin{figure}
\centering
\includegraphics[width=12cm,height=6cm]{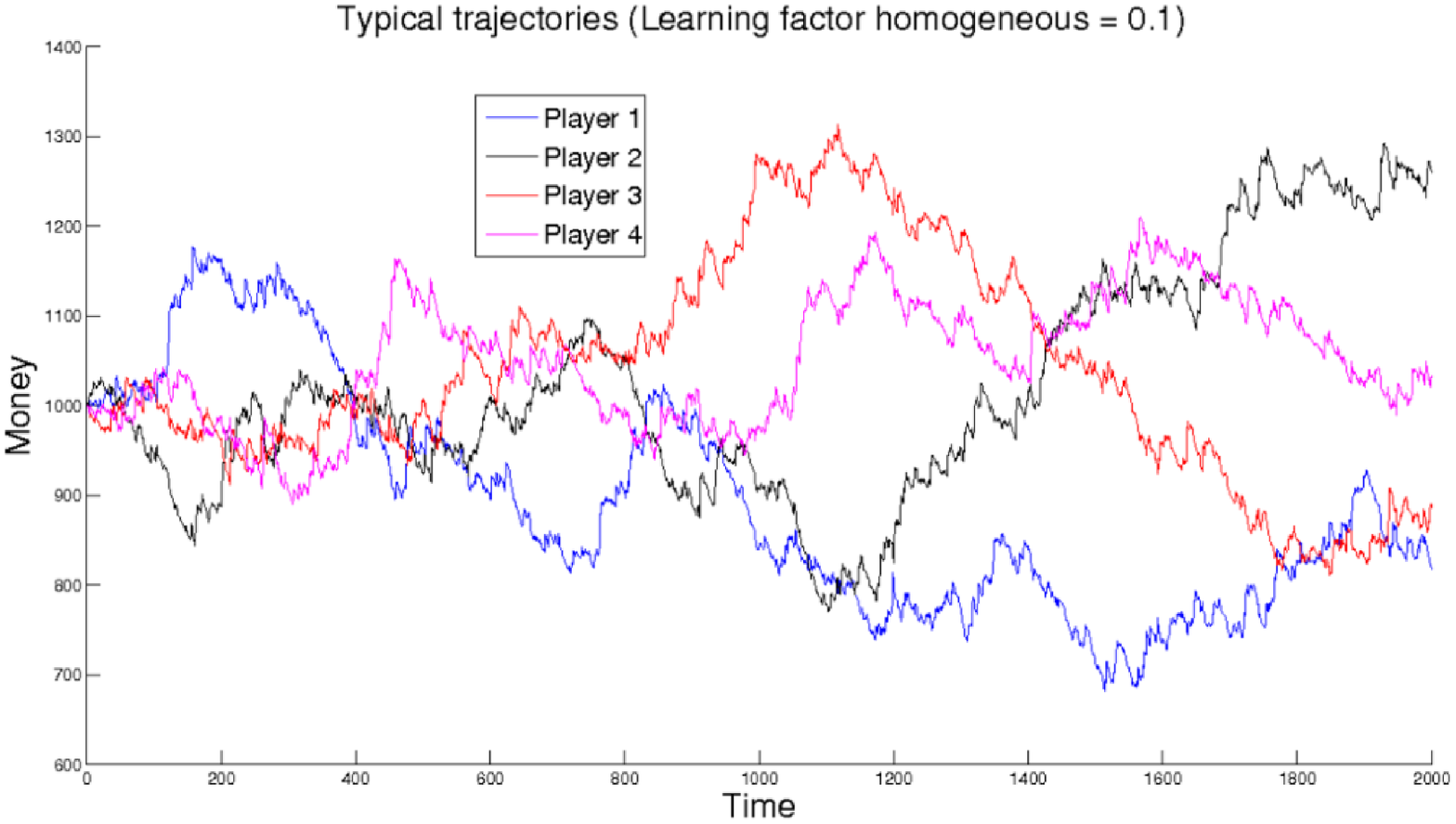}
\caption{}\label{fig:typtraj2}
\end{figure}

\begin{figure}
\centering
\includegraphics[width=12cm,height=6cm]{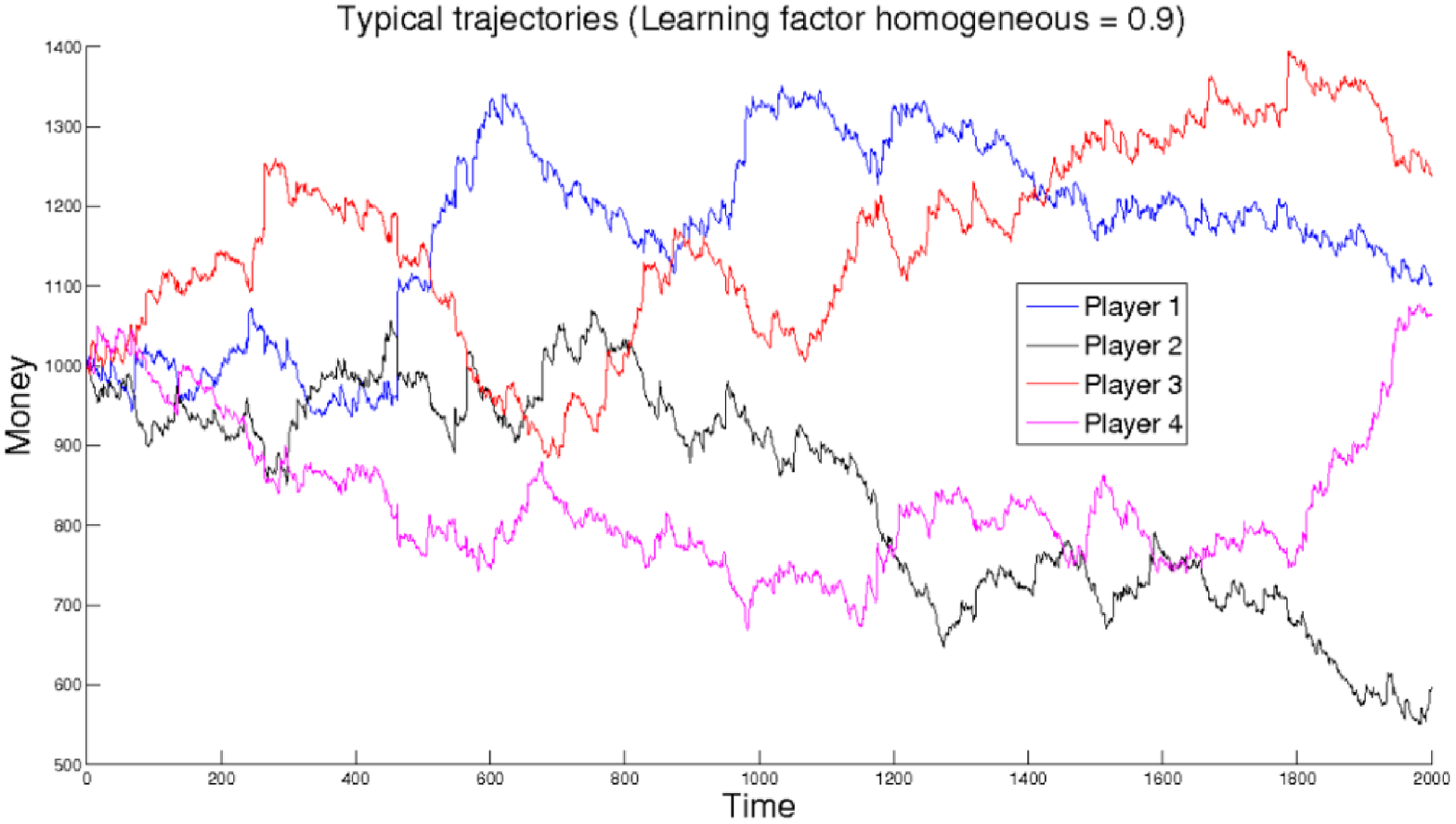}
\caption{}\label{fig:typtraj3}
\end{figure}

Through the variation of the LF value for each player some characteristics of the system appears unvaried while other aspects result changed, thus testifying an effect of the learning factor also in an homogeneous system.
More in depth the figures \ref{fig:typtraj2} and \ref{fig:typtraj3} allows a comparison between the trajectories of the agent respectively in the case of  $LF=0.1$ and  $LF=0.9$. At a first inspection of the figures it seems that the regime of the fluctuations of the system grows faster in the second condition that it does in the first one.

\begin{figure}
\centering
\includegraphics[width=12cm,height=6cm]{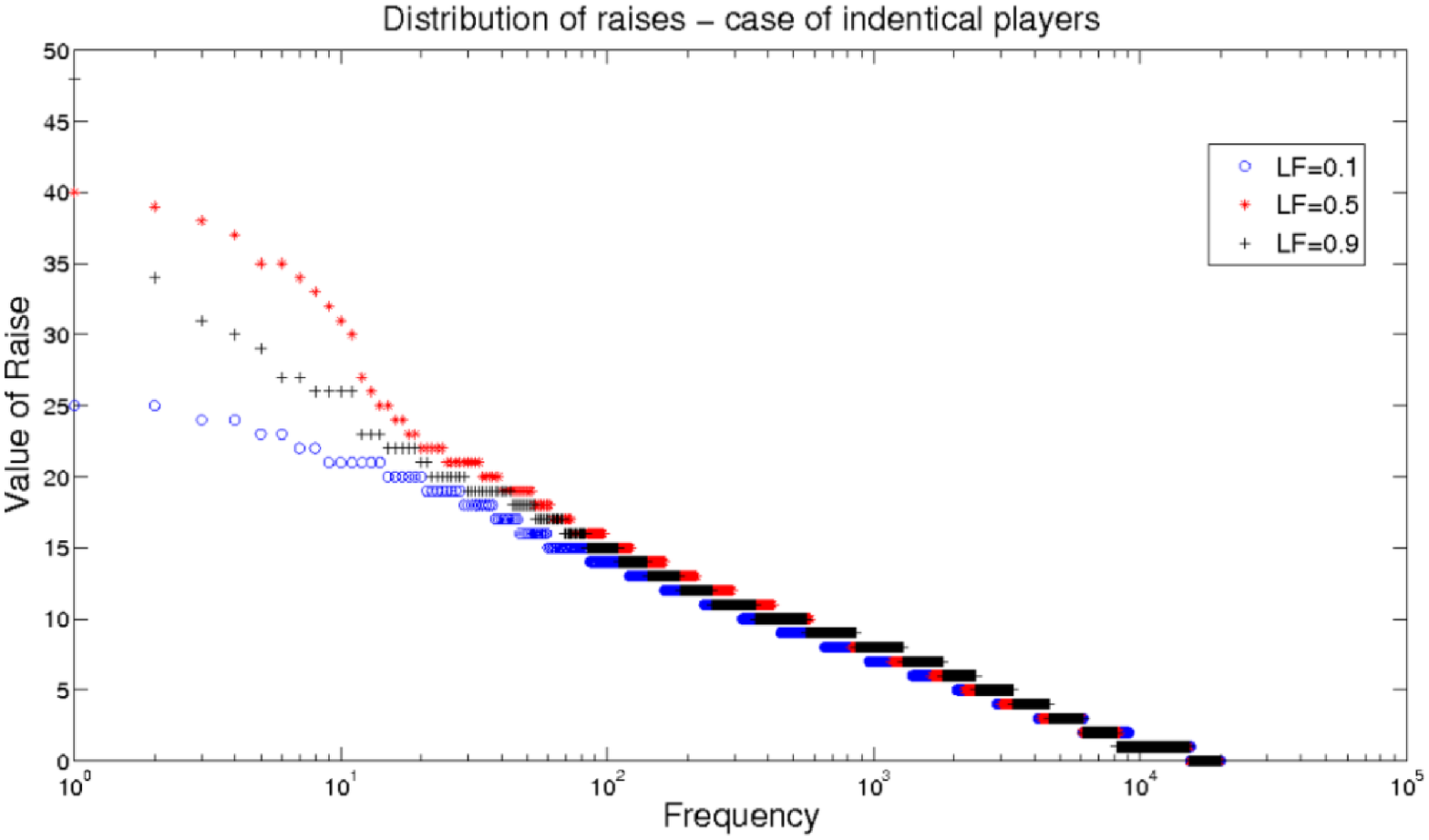}
\caption{}\label{fig:Omodist}
\end{figure}

Furthermore the analysis of the distributions of the players' raises, peculiar to the three different homogeneous systems in object, shows other interesting aspects. First of all these adopt an exponential trend, as the one shown by \ref{fig:Omodist}. Moreover taking into consideration the analysis of  what we may interpret as the effects of finite size that characterize the right part of all the three distributions,  an interesting aspect is represented by the fact that the most considerable raises show up in the system with an intermediate LF ($0.5$).

\begin{figure}
\centering
\includegraphics[width=15cm,height=13cm]{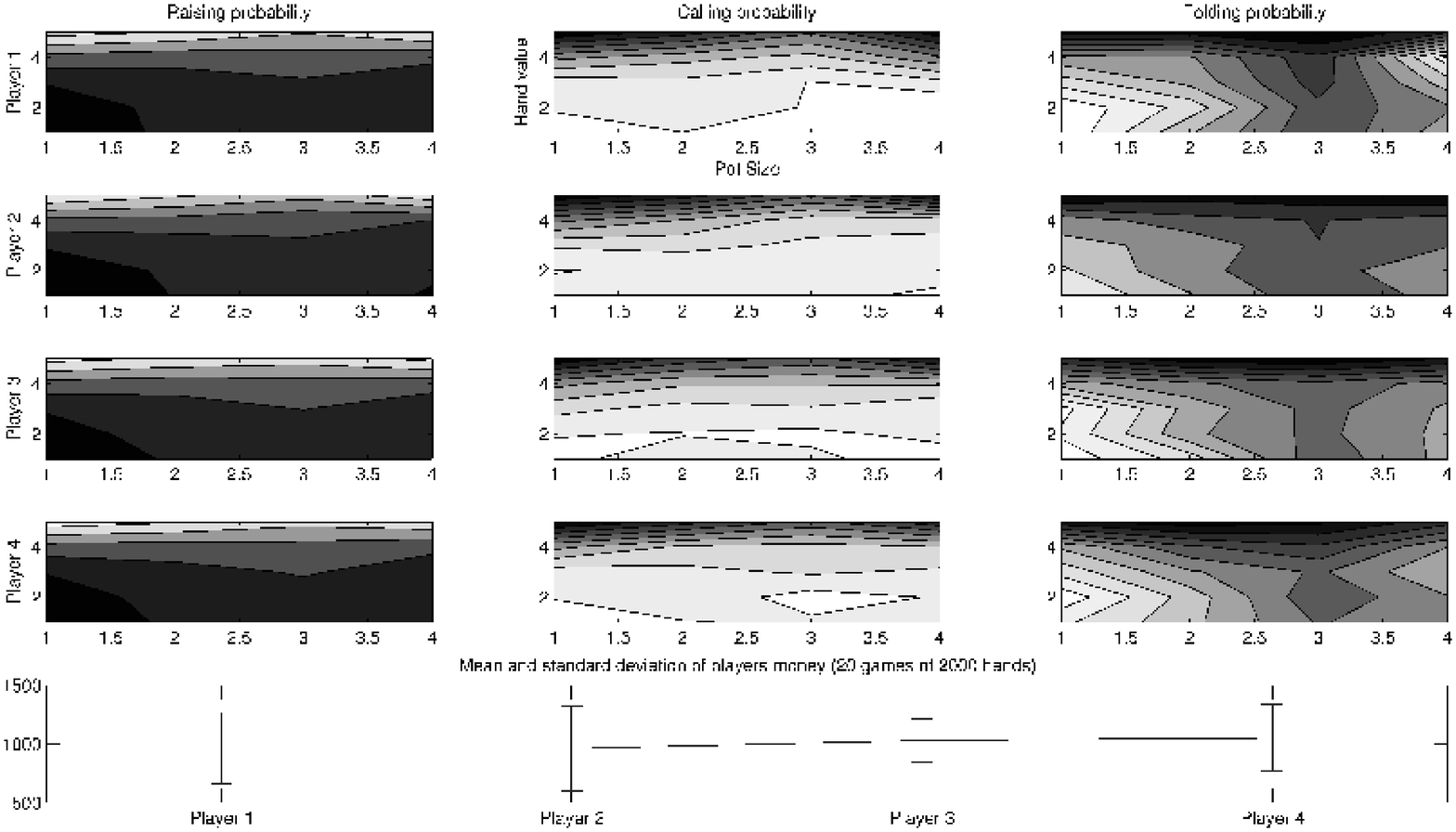}
\caption{For each player the three matrices which define the player's stable strategy are obtained considering the mean among 20 different games of 2000 hands. In the bottom of the figure a mean and standard deviation of player's money in the final state is also presented}\label{fig:Multi01}
\end{figure}

\begin{figure}
\centering
\includegraphics[width=15cm,height=13cm]{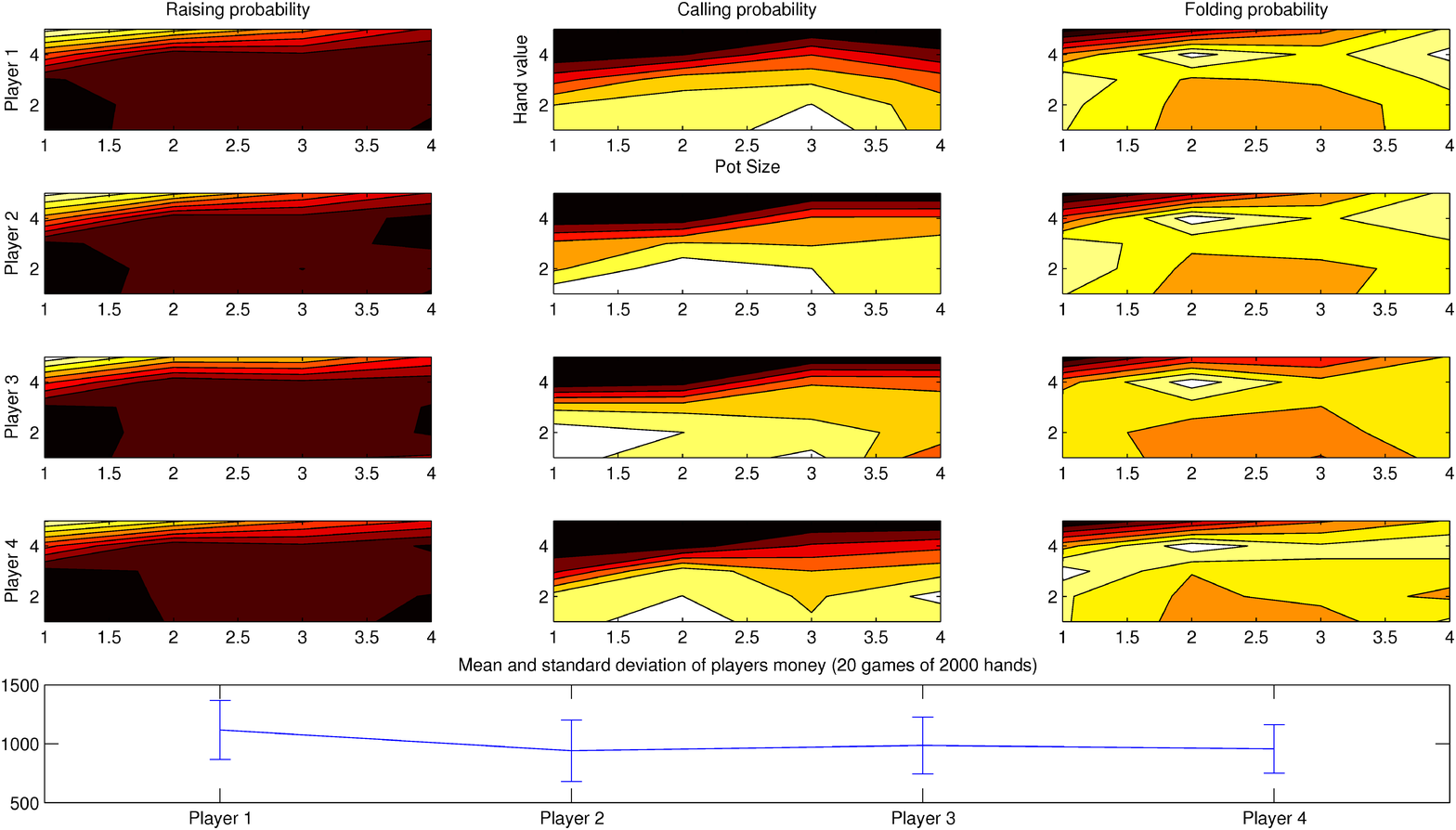}
\caption{For each player the three matrices which define the player's stable strategy are obtained considering the mean among 20 different games of 2000 hands. In the bottom of the figure a mean and standard deviation of player's money in the final state is also presented}\label{fig:Multi09}
\end{figure}

As a matter of fact, taking into consideration also the stable strategies which characterize the agents in each one of the three experimental conditions it is possible to notice how the most daring strategy (e.g. the one characterized by the highest tendency both towards raise and bluff) is right the one developed by the agents in a system with intermediate LF, while the most conservative one seems to be developed in a system where all the agents are characterized by an high LF (\ref{fig:Multi01} \& \ref{fig:Multi09}).

\subsubsection{LF-Heterogeneous system}

In this second condition all but one agent were characterized by the same LF,  fixed in a consevative way at 0.5. The 'strange' player's LF instead has been assumed as control parameter and has been varied consequently. Also for this condition at the beginning of each game all elements of strategie's matrices were fixed at 1/3.

\begin{figure}
\centering
\includegraphics[width=12cm,height=8cm]{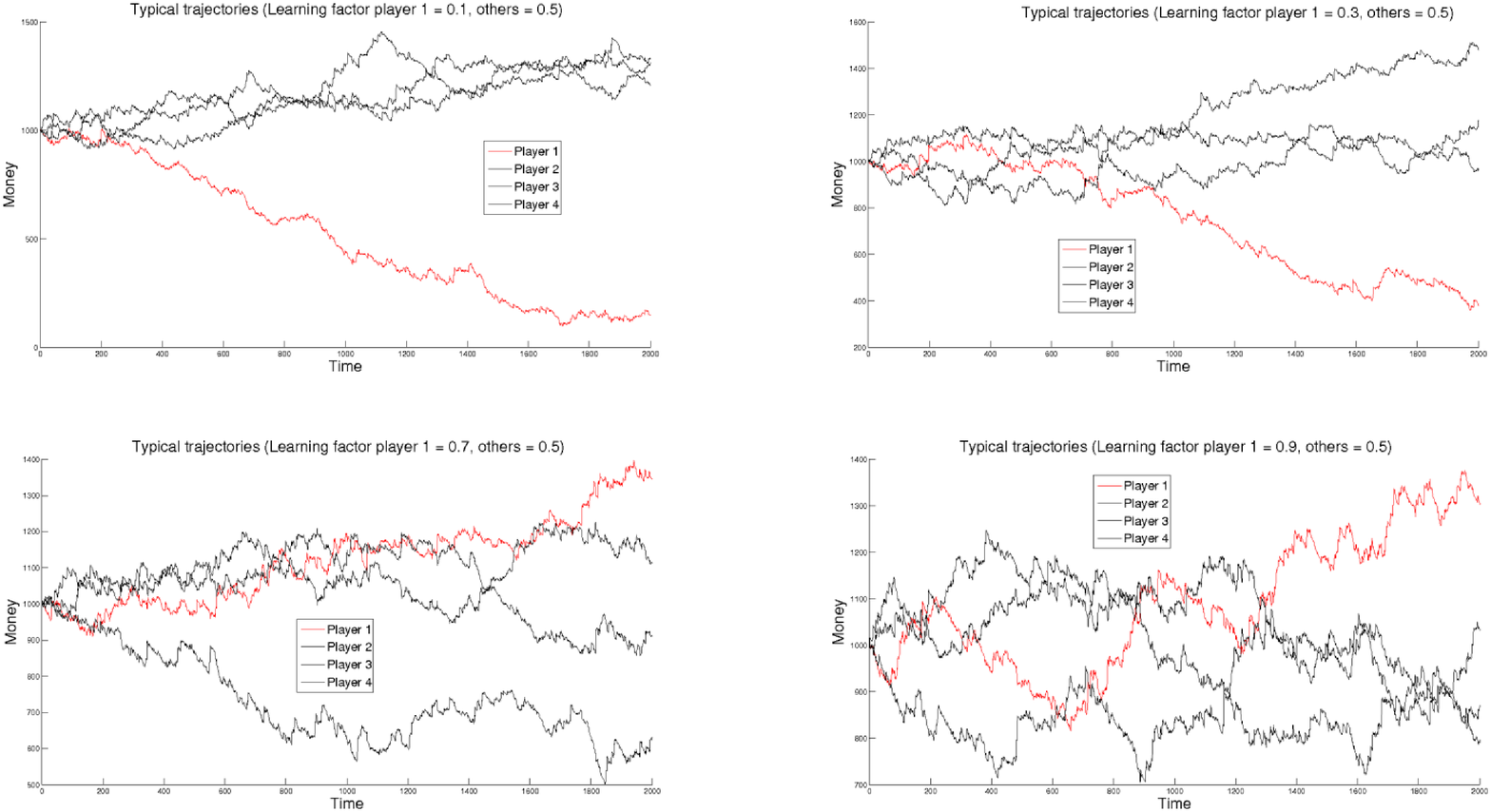}
\caption{}\label{fig:Multitraj}
\end{figure}

The visual inspection of different single game typical trajectories, relative to the different conditions, point out a clear effect of learning factor \ref{fig:Multitraj}).

\begin{figure}
\centering
\includegraphics[width=12cm,height=8cm]{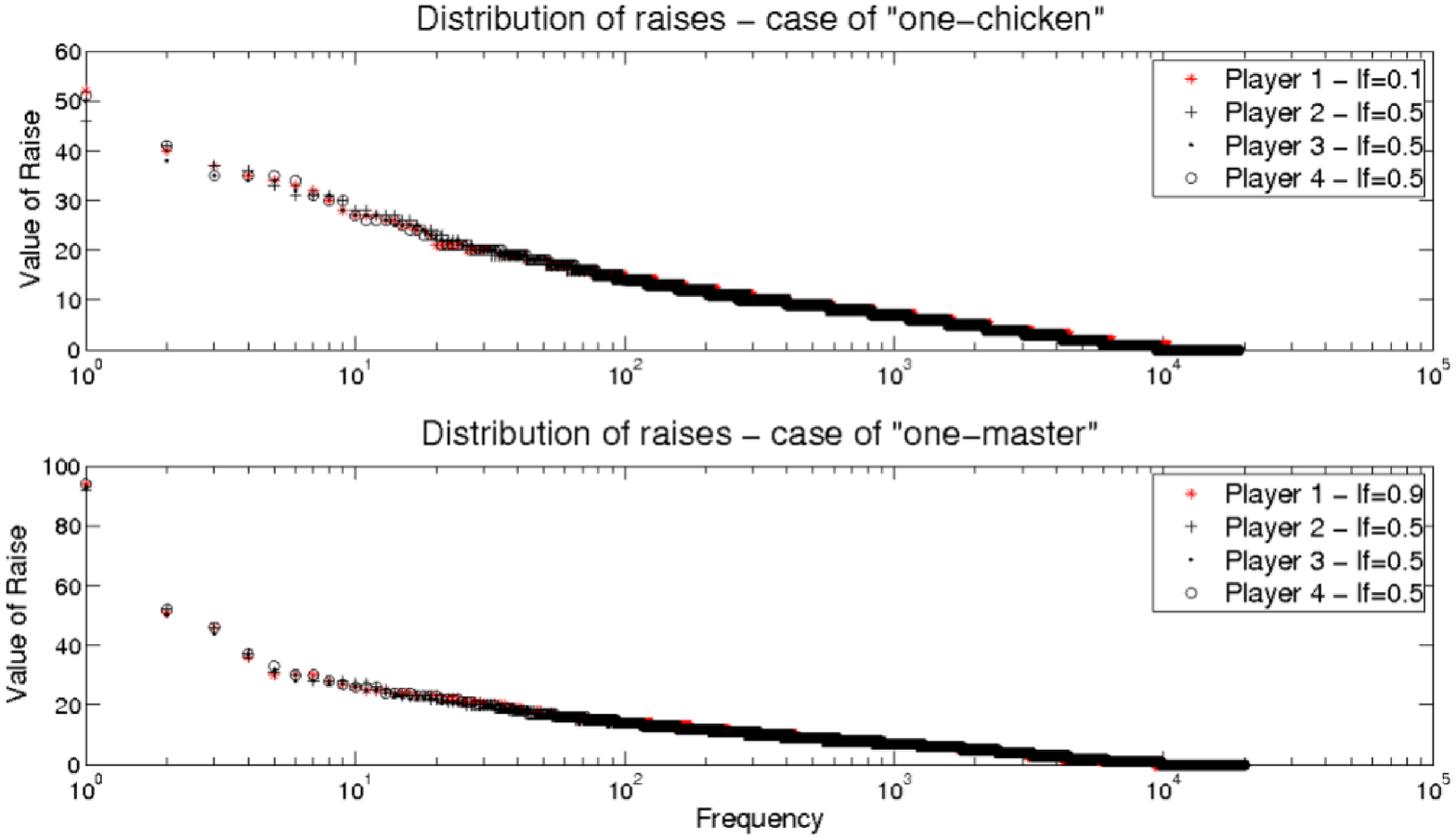}
\caption{}\label{fig:Eterodist}
\end{figure}

Furthermore the trend of the distribution of the raises for the single players, in  the two extreme conditions defined respectively \textit{one-chicken} and \textit{one-master} cases shows how the intermediate Learning Factor seems to positively correlate with the maximum observed dimension of the raises (figure \ref{fig:Eterodist}). 

Moving on to the analysis of the strategies developed by the agents in the two above mentioned extreme cases , and to the evaluation of their fitness other interesting phenomena emerge.

\begin{figure}
\centering
\includegraphics[width=15cm,height=13cm]{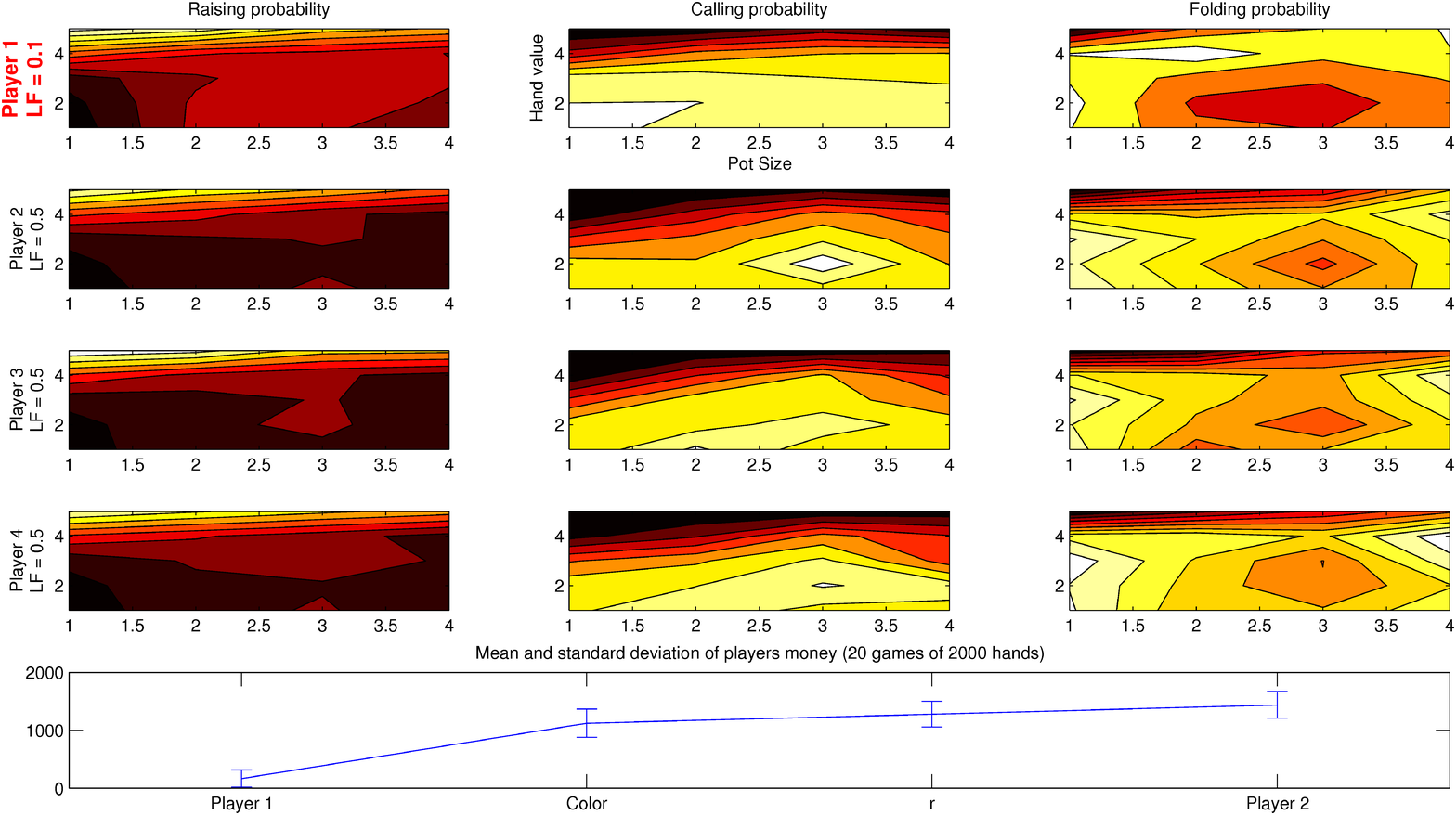}
\caption{For each player the three matrices which define the player's stable strategy are obtained considering the mean among 20 different games of 2000 hands. In the bottom of the figure a mean and standard deviation of player's money in the final state is also presented}\label{fig:Multy01Vs05}
\end{figure}

Also in the case of the heterogeneous system both rational strategies and tendencies to bluff seem to emerge in all the players and in all the conditions. Yet an accurate analysis shows how in the case defined \textit{once chicken case} the agent characterized by the LF of inferior to the average tends to have the most daring strategy \ref{fig:Multy01Vs05} even if it frequently turns out to be the one  which loses the most. This notwithstanding, the other agents develop very similar strategies and also similar to the ones of the homogeneous condition and characterized by more conservative tactics with respect to the first agent as far as the raise is concerned. Yet such strategies are characterized by a high probability of seeing the opponent's raise, but only when the pot is conspicuous and also for  average low values of its own hand.  This apparently counter-inductive tendency may be seen as another way of bluffing. This strategy probably represents the best adaptive answer that a player, with much money, may adopt with an adversary who, despite its  smaller amount of money, bluffs a lot.  Furthermore, as it is well known to every poker player, the tendency to bluff tends to increase at the decreasing of the size of the available money, even if it should not behave like that.

\begin{figure}
\centering
\includegraphics[width=15cm,height=13cm]{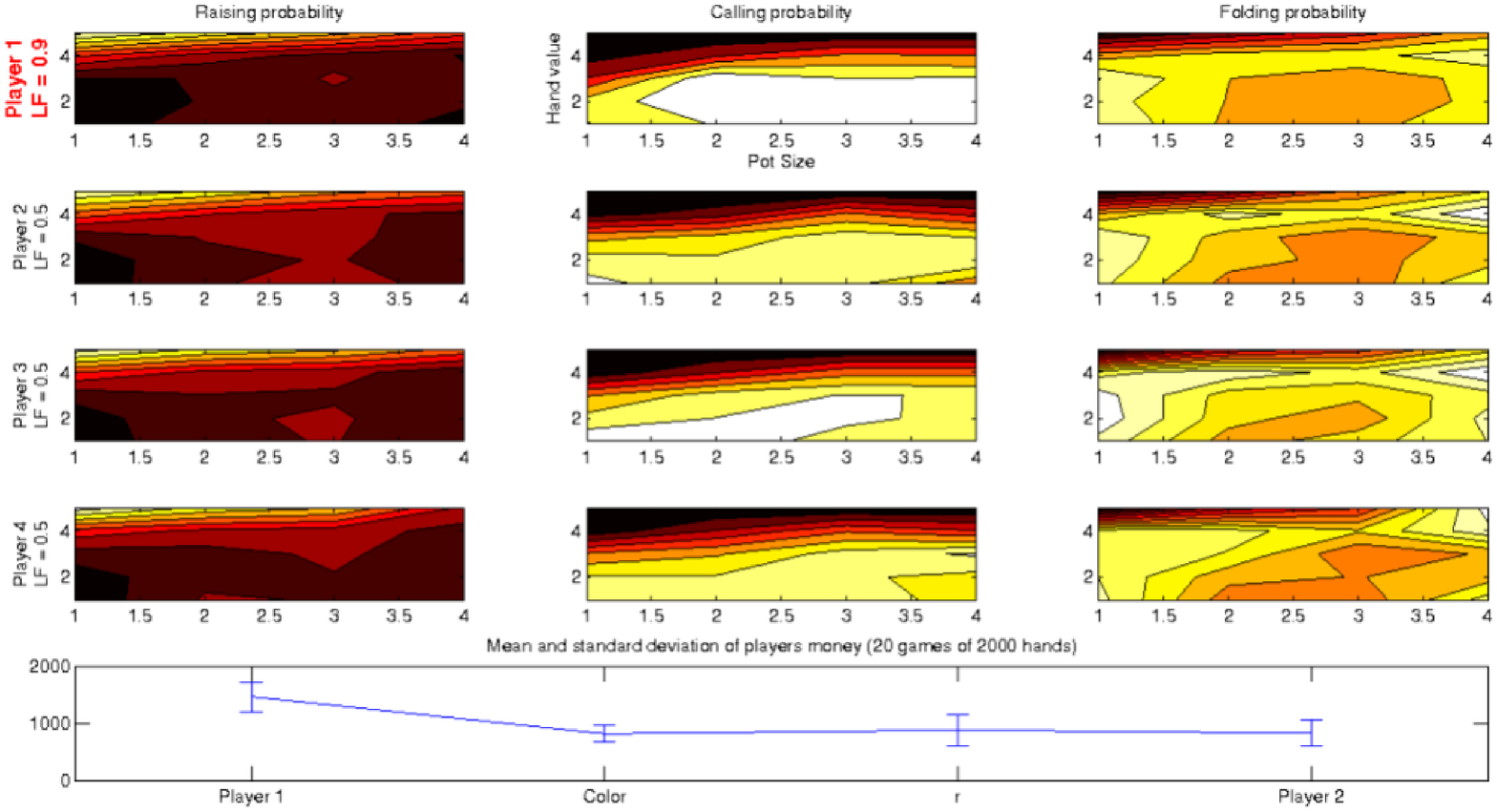}
\caption{For each player the three matrices which define the player's stable strategy are obtained considering the mean among 20 different games of 2000 hands. In the bottom of the figure a mean and standard deviation of player's money in the final state is also presented}\label{fig:Multy09Vs05}
\end{figure}

Instead in the case defined as \textit{one master case} the fitness of strange agent seems to be definitely much better, as shown by the graphic in the lower part of figure \ref{fig:Multy09Vs05}). In this case as well it is the agent characterized by the lowest LF that seem to develop the most daring strategy, and also in this case the agent with the highest LF develops what we would define an apparent insane inclination to call, exploiting in this case as well its higher tolerance to the fluctuations and through the development of this strategy mostly in answer to the dominant strategy of its own adversaries.

\begin{figure}
\centering
\includegraphics[width=12cm,height=6cm]{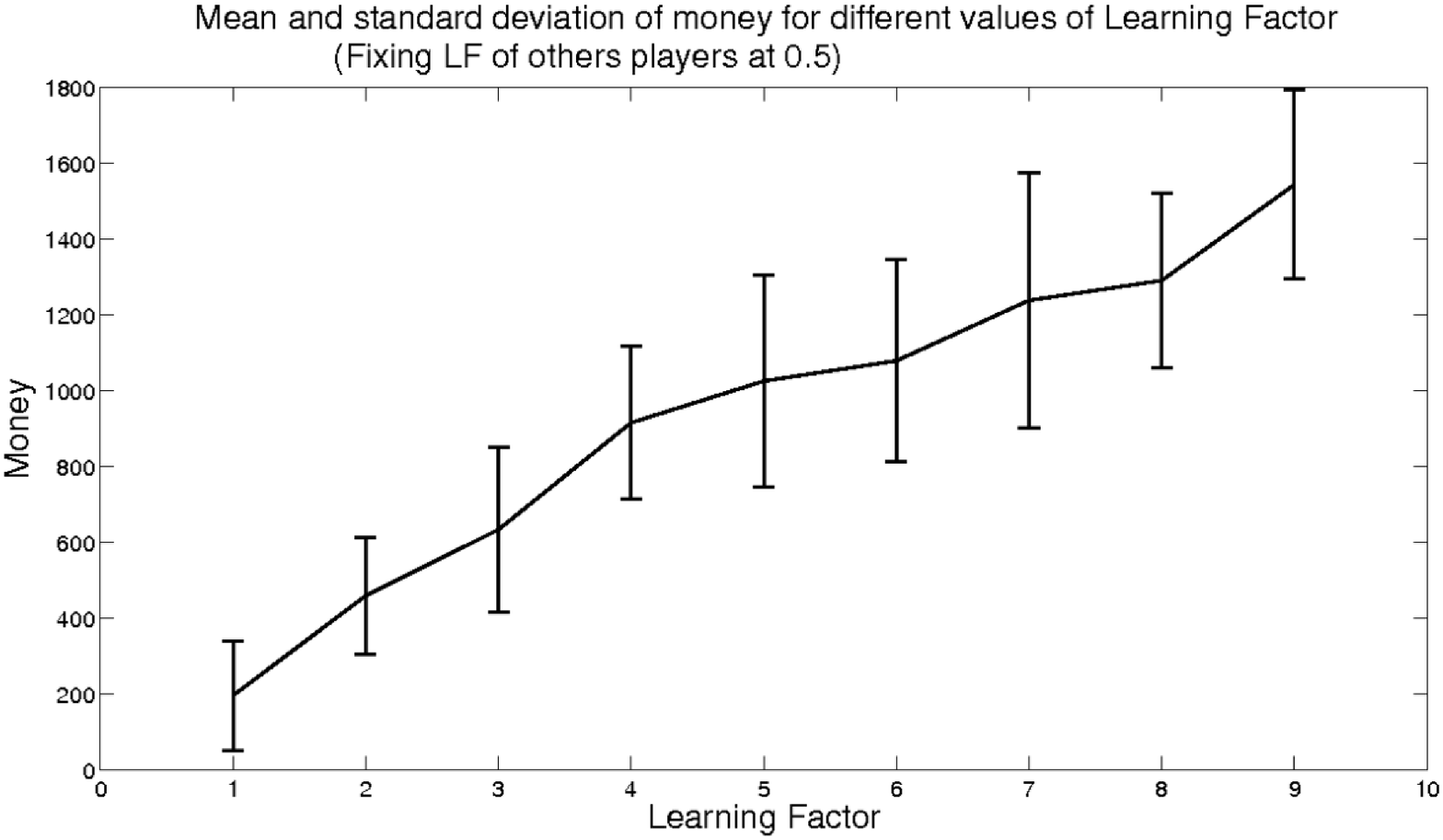}
\caption{}\label{fig:learning}
\end{figure}

Finally in order to determine the relationship between the LF and the fitness of the agents it was taken into consideration a system where three agents out of four were characterized by an intermediate LF ($0.5$) by making the last player adopt different LF values and by eventually studying the relationship  between this last  and the average of the money in the space of $20$ simulations for each LF value. 
The \ref{fig:learning})  stresses how the fitness (represented by the player's average gain) tends to increase with the increasing of its learning speed moving on from a negative fitness, for LF Values inferior than the others player's ones, to a positive fitness for superior LF values. Moreover even complete heterogeneity conditions of the system, i.e. with four players characterized by four different LF, tend to follow dynamics which are similar to those so far noticed, that is to say they both develop the bluff, and they differ because of a positive correlation between fitness and learning factor.

\section{Second part: the two agent model}

The model we are going to define and analyse in this section is probably one of the greatest simplifications possible of
a game of chance with imperfect information.

Here we have two players: at the beginning of each hand of the game, they put one coin as the entry pot. Then, they
pick a "card" from a pack: each card has an integer value between $0$ and $N-1$ ({\it i.e.\rm} there are $N$ cards
overall). At this point, according to the value of their card, the players decide to call or instead to fold. If both
players call, they put another coin in the pot, and who holds the highest card wins: the winner gets the entire pot
(four coins). If one of the players folds, the "caller" wins and gets the entry pot (two coins). Finally, if nobody
calls, both players take back the coin they had put as entry pot.
Mathematically, when the player $i$ holds the card of value $n$, decides to call according to the probability
distribution $P_i(n)$, with of course $i=1,2$ and $n\in\{0,1,\dots,N-1\}$. After every hand both players update their
strategy. More precisely, if one folds, nothing happens; if the agent $i$, thanks to the card $n$, calls and wins (because
he has the highest card or because the opponent folds), the probability that he calls holding the card $n$ will change
in this way:

\begin{equation}
\label{winupd}
P_i(n)\longrightarrow P_i(n)+\mu_i[1-P_i(n)]
\end{equation}

Analogously, if the agent $i$ loses (that is, if he calls but the opponent has a higher card than him), the
probability $P_i(n)$ will change instead in the following way:

\begin{equation}
\label{loseupd}
P_i(n)\longrightarrow\mu_i P_i(n)
\end{equation}

In Equations (\ref{winupd}) and (\ref{loseupd}) the coefficient $\mu_i$ is again the learning factor, which can also be
seen as a sort of risk propensity of the player $i$. The LFs of the players are set at the beginning of the game,
and will never change. Moreover, it can assume a value between $0$ (no risk propensity at all), and $1$ (maximum
risk propensity possible): actually, a player with $\mu_i=0$ and the card $n$ does not increase $P_i(n)$ even though
he wins and sets $P_i(n)=0$ as soon as he loses; instead, with $\mu_i=1$ he sets $P_i(n)=1$ when he wins but does not
decrease $P_i(n)$ if he loses.
Finally, it is easy to notice that Equations \ref{winupd} and \ref{loseupd} ensure that
$P_i(n)$ will always stay in the interval $[0,1]$.

\subsection{Numerical results}

In this subsection we will present the most remarkable results of the simulations of the simple model defined above.

First of all, for simplicity we set the LF $\mu_1$ of the "player $1$" equal to $0.5$, and then we checked the
dynamics by varying $\mu_2$: actually it is the difference between the LFs which essentially determines the
main features of the dynamics, as we saw in several simulations. In particular, we can distinguish three cases:
$\mu_2<\mu_1$, $\mu_2=\mu_1$, and $\mu_2>\mu_1$.

\subsubsection{Case $\mu_2<\mu_1$}

\begin{figure}[h]
\begin{center}
\includegraphics[angle=0,width=8cm,clip]{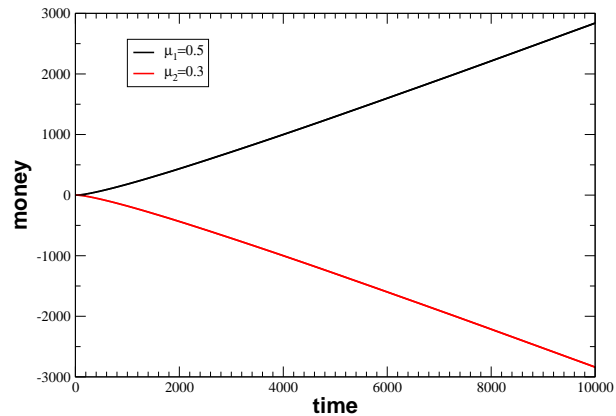}
\end{center}
\caption{
Behaviour of the money won (or lost) by the two players versus time, {\it i.e.\rm} versus the hands played, averaged
over $10 ^4$ iterated matches. The LFs of the players are here $\mu_1=0.5$ and $\mu_2=0.3$.}
\label{case1figA}
\end{figure}

\begin{figure}[h]
\begin{center}
\includegraphics[angle=0,width=8cm,clip]{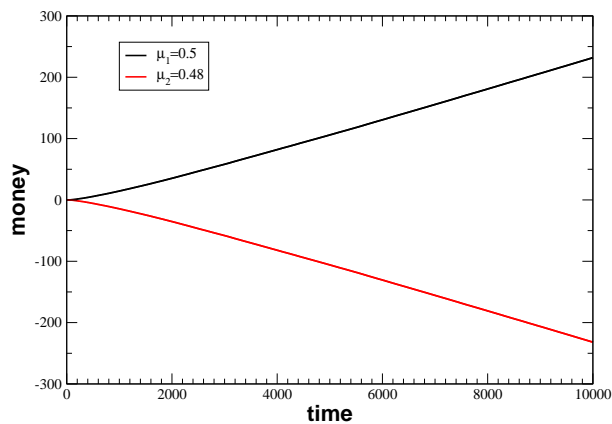}
\end{center}
\caption{
Behaviour of the money won (or lost) by the two players versus time, {\it i.e.\rm} versus the hands played, averaged
over $10 ^5$ iterated matches. The LFs of the players are here $\mu_1=0.5$ and $\mu_2=0.48$.}
\label{case1figB}
\end{figure}

In figures \ref{case1figA} and \ref{case1figB} it is shown the behaviour of the money of both players as a function
of time, where the time unit is a single hand of the game: we set $N=25$, $\mu_2=0.3$ and $\mu_2=0.48$ respectively,
and the results are averaged over $10^4$ and $10^5$ iterations, respectively; in both cases, we let the agents play
$10^4$ hands. For simplicity, we considered players with an infinite amount of money available, and we gauged to zero
the initial amount. Additionally, the initial calling distributions $P_i(n)$ are picked randomly for each $n$.


As it can be seen, the first player, with higher LF, wins over his opponent, with smaller LF, and the money gained by
player $1$ increases with time. Moreover, the smaller is $\mu_2$, the faster and bigger are the winnings of player $1$.
Previous figures show that on average the player with bigger risk propensity finally overwhelms the other one, and
this means that in a single match the most risk-inclined player has a bigger probability to win, and such probability
increases as the difference $\mu_1-\mu_2$ increases in its turn.
The fact that risking is convenient gets confirmed in the next figures, where the final calling distributions for both
players are depicted.

\begin{figure}[h]
\begin{center}
\includegraphics[angle=0,width=8cm,clip]{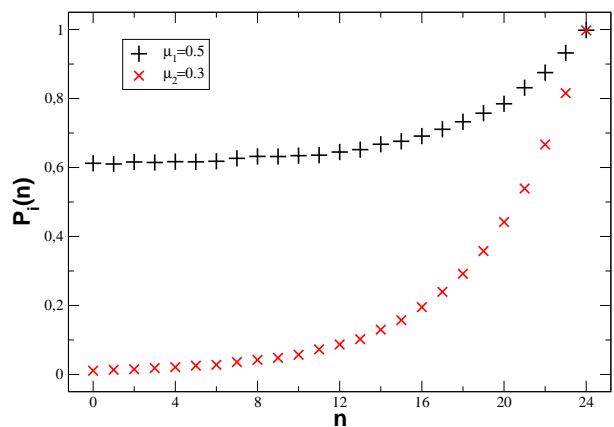}
\end{center}
\caption{
Calling probabilities as functions of the card value $n$ for both players. Black symbols: player 1 ($\mu_1=0.5$);
red symbols: player 2 ($\mu_2=0.3$). Data took after $10^4$ hands of the game, and averaged over $10^4$ iterations.
Random initial distribution for every $P_i(n)$.}
\label{case1figC}
\end{figure}
\begin{figure}[h]
\begin{center}
\includegraphics[angle=0,width=8cm,clip]{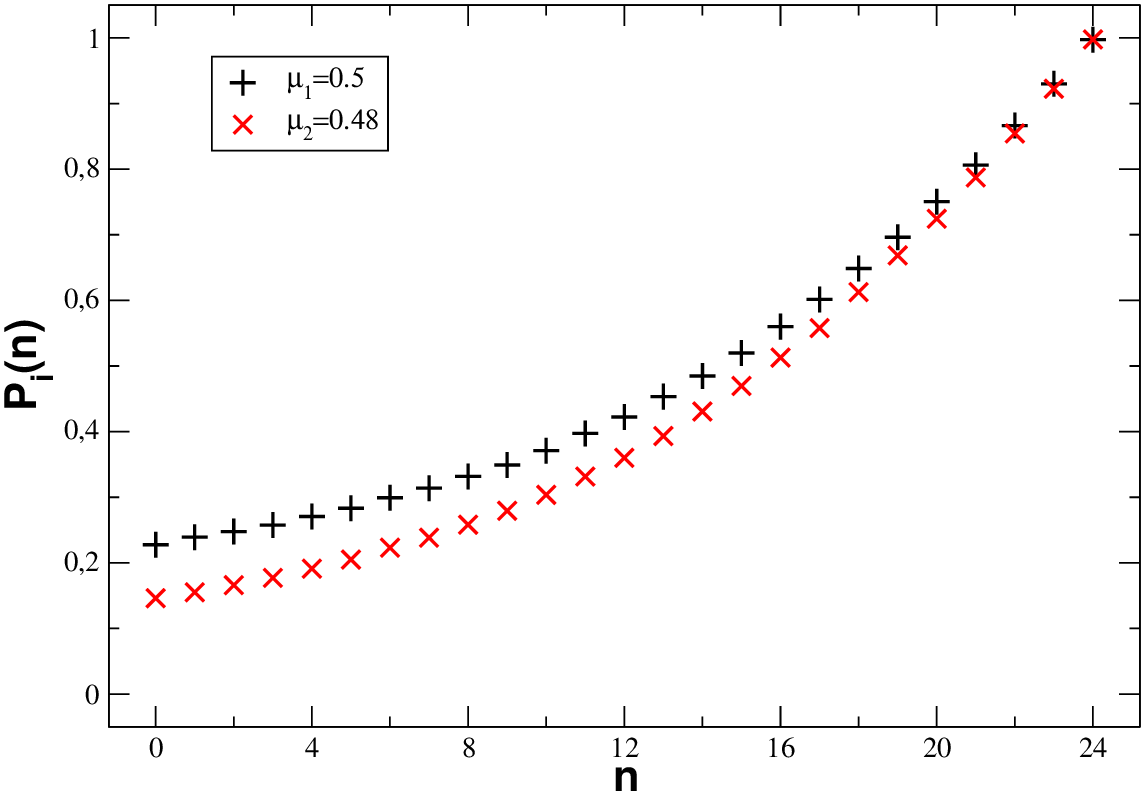}
\end{center}
\caption{
Calling probabilities as functions of the card value $n$ for both players. Black symbols: player 1 ($\mu_1=0.5$);
red symbols: player 2 ($\mu_2=0.48$). Data took after $10^4$ hands of the game, and averaged over $10^5$ iterations.
Random initial distribution for every $P_i(n)$.}
\label{case1figD}
\end{figure}

As we can see, in both cases the winner is characterized by higher calling probabilities than his opponent's ones
for every value of $n$. Moreover, we have $P_i(n=0)>0$, which is the most explicit evidence of the emergence
of bluffing.

\subsubsection{Case $\mu_2=\mu_1$}

\begin{figure}[h]
\begin{center}
\includegraphics[angle=0,width=8cm,clip]{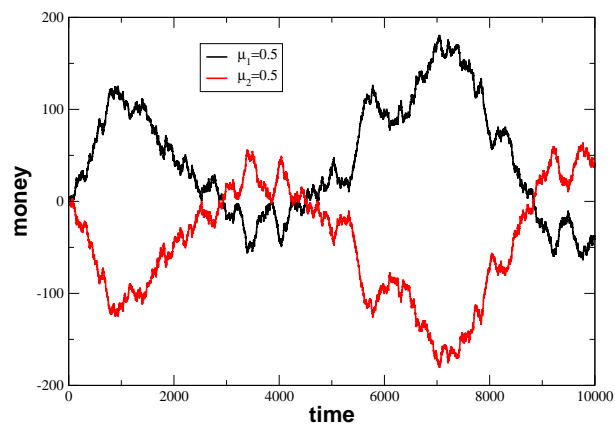}
\end{center}
\caption{
Behaviour of the money won (or lost) by the two players versus time, {\it i.e.\rm} versus the hands played, throughout
a single match. The LFs of the players are here $\mu_1=\mu_2=0.5$.}
\label{case1figE}
\end{figure}

In this case, both players have the same winning probability, as it is well shown in figure \ref{case1figE}: indeed,
having the same LF, they have also exactly the same behaviours so that in a single match nobody is able to overwhelm
definitively the opponent. 

Naturally, it is also easy to forecast the behaviour of the final calling probabilities of the agents: they will be
equal, and with $P_i(n=0)>0$ for both $i=1$ and $i=2$.

\begin{figure}[h]
\begin{center}
\includegraphics[angle=0,width=8cm,clip]{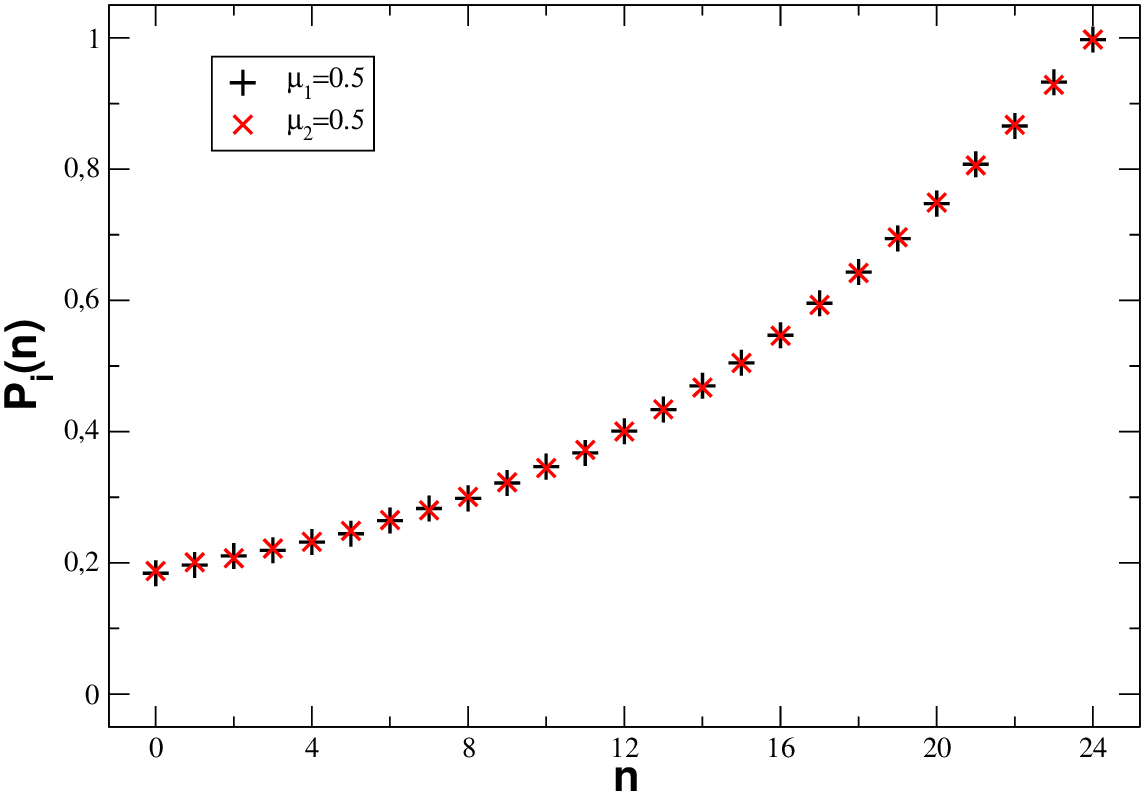}
\end{center}
\caption{
Calling probabilities as functions of the card value $n$ for both players. Black symbols: player 1 ($\mu_1=0.5$);
red symbols: player 2 ($\mu_2=0.5=\mu_1$). Data took after $10^4$ hands of the game, and averaged over $2.5\cdot10^4$ iterations.
Random initial distribution for every $P_i(n)$.}
\label{case1figF}
\end{figure}

\ 

\subsubsection{Case $\mu_2>\mu_1$}

\begin{figure}[h]
\begin{center}
\includegraphics[angle=0,width=8cm,clip]{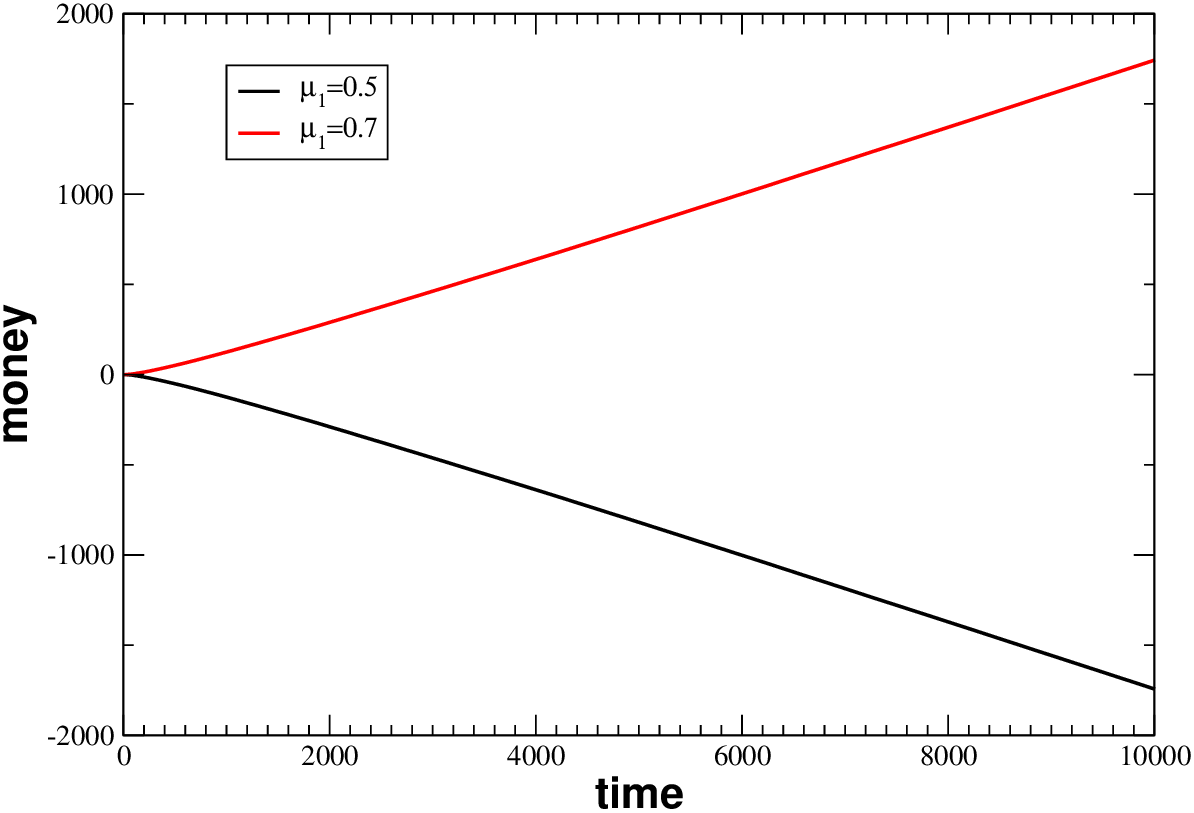}
\end{center}
\caption{
Behaviour of the money won (or lost) by the two players versus time, {\it i.e.\rm} versus the hands played, averaged
over $10 ^4$ iterated matches. The LFs of the players are here $\mu_1=0.5$ and $\mu_2=0.7$.}
\label{case1figG}
\end{figure}

\begin{figure}[h]
\begin{center}
\includegraphics[angle=0,width=8cm,clip]{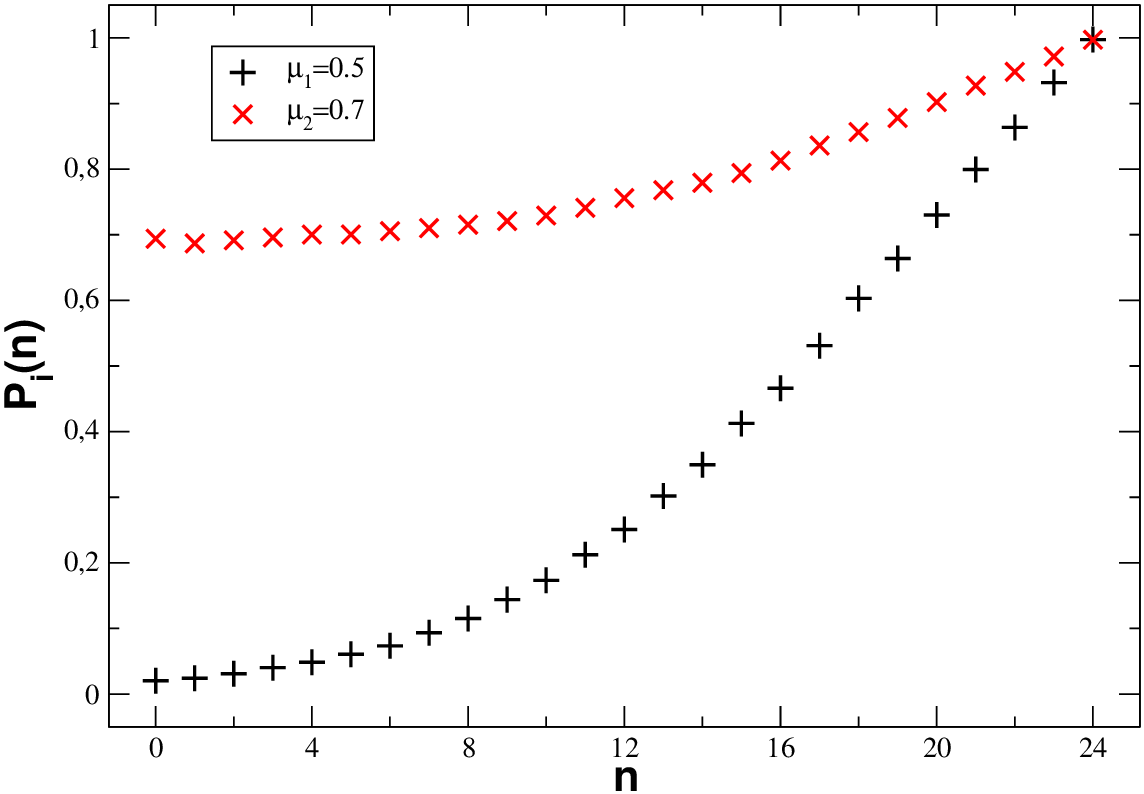}
\end{center}
\caption{
Calling probabilities as functions of the card value $n$ for both players. Black symbols: player 1 ($\mu_1=0.5$);
red symbols: player 2 ($\mu_2=0.7$). Data took after $10^4$ hands of the game, and averaged over $10^4$ iterations.
Random initial distribution for every $P_i(n)$.}
\label{case1figH}
\end{figure}

In this case the results are qualitatively equal to the ones of case $\mu_2>\mu_1$, only now it is player $2$ which
defeats player $1$, as shown in figures \ref{case1figG} and \ref{case1figH}.


\subsection{Discussion}

The first conclusion we can obtain just from the numerical results is that in this game bluffing emerges necessarily
as rational strategy.
Moreover, the player who more bluffs finally wins. This can be easily understood watching figures \ref{case1figC},
\ref{case1figD}, \ref{case1figF}, and \ref{case1figH}. Indeed, while in general the calling probabilities of both
players tend to have the same value for $n\rightarrow N-1$, for small $n$ the player with higher LF, that is with
higer risk propensity, bluffs much more than his opponent: this means that even holding a poor card, the "risk-lover" will call and unless his opponent has a very good point, will get the entry pot.

It is possible to formalize such considerations by writing the equations of the dynamics for the model at stake.
Neglecting fluctuations, the "mean-field" equation ruling the money $M_1$ won (or lost) by the first player is

\begin{eqnarray}
M_1(t+1) & = & M_1(t)+\sum_{n_1=0}^{N-1}\frac{1}{N}\left[(1-\hat\Pi^1_2)P_1(n_1;t)-\hat\Pi^1_2(1-P_1(n_1;t))+\right. \nonumber \\
	& + &\left. 2P_1(n_1;t)\hat\Pi^1_2(\mathcal{P}^1_2-\mathcal{P}^2_1)\right]
\label{MF1}
\end{eqnarray}

where $\mathcal{P}^1_2=\Pr(n_1>n_2)$ is the probability that the card $n_1$ held by player $1$ is haigher than the card
$n_2$ held by player $2$; analogously it is $\mathcal{P}^2_1=\Pr(n_2>n_1)$. On the other hand $\hat\Pi^1_2$ is an operator
defined as follows

\[
\hat\Pi^1_2\cdot X=\sum_{n_2=0}^{N-1}\left[\frac{P_2(n_2;t)}{N-1}(1-\delta_{n_2,n_1}) X\right]
\]

and represents the probability that player $2$ calls from the point of view of player $1$. Equation (\ref{MF1}) can be
rewritten as a differential equation in time, which can assume the form

\begin{equation}
\dot M_1(t)=\left[1-2\gamma_2(t)\right]\omega_1(t)-\left[1-2\gamma_1(t)\right]\omega_2(t)
\label{MF2}
\end{equation}

with

\begin{equation}
\omega_i(t)=\frac{1}{N}\sum_{n=0}^{N-1}P_i(n;t)\ \ \ \ \ \ \ \ \ \ \ \ \ i=1,2
\label{MF3}
\end{equation}

and

\begin{equation}
\gamma_i(t)=\frac{1}{(N-1)^2}\sum_{n=0}^{N-1}nP_i(n;t)\ \ \ \ \ \ \ \ \ \ \ \ \ i=1,2
\label{MF4}
\end{equation}

Since this is a zero-sum game, second player's money will be obtained by the relation $M_2(t)=-M_1(t)$.

Finally, the relation giving the temporal behaviour of the calling distributions $P_1(n;t)$ of player $1$ (being the
one of the opponent of analogous form) is

\begin{eqnarray}
\dot{P}_1(n_1;t) & = & \frac{1}{N}P_1(n_1;t)\left[(1-\hat\Pi^1_2+\hat\Pi^1_2\mathcal{P}^1_2)[P_1(n_1;t)+ \right.\nonumber\\
	& + & \left. \mu_1(1-P_1(n_1;t))] +\mu_1P_1(m_1;t)\hat\Pi^1_2\mathcal{P}^2_1\right]+\nonumber\\
	& + & \frac{1-P_1(n_1;t)}{N}P_1(n_1;t)-\frac{P_1(n_1;t)}{N}
\label{MF5}
\end{eqnarray}

Now, equations (\ref{MF2}), (\ref{MF3}), (\ref{MF4}), and (\ref{MF5}) are rather complicated, but some features
of them can be determined without an explicit solution. Actually, it is straightforward to understand that we have

\[
P_1(n;t)=P_2(n;t)\Longrightarrow\omega_1(t)=\omega_2(t),\ \gamma_1(t)=\gamma_2(t)\Longrightarrow\dot{M}_1=\dot{M}_2=0
\]

Now, since in our simulations we always started from the same initial $P_i(n)\ \forall i,n$, and from
$M_1(0)=M_2(0)=0$, this implies that for $\mu_1=\mu_2$ both player must have on average the same calling
distributions and then they should not gain nor lose money, apart from fluctuations: this is exactly what
we found in figures \ref{case1figE} and \ref{case1figF}. It can also be shown that, if $\mu_1>\mu_2$, then
we will have soon $P_1(n)\geq P_2(n)\ \forall n$ (with the equality holding only for $n=N-1$), so that,
equations (\ref{MF3}) and (\ref{MF4}) allow us to get $\dot{M}_1>0$, that is the victory of player1, as shown
in figures from \ref{case1figA} to \ref{case1figD}. Obviously, the opposite situation takes place for $\mu_1<\mu_2$
(as shown in figures \ref{case1figG} and \ref{case1figH}).

\section{Discussion and conclusions}

In this work we have developped a model to describe cognitive processes dynamics. First of all, cognitive sciences theories gave us the main ingredients to write down simple poker models. Secondly, by an agent based approach we investigated the role of some critical features of the interaction between agents and environment for the evolution of rational strategies.
The only feature taken into account to characterizes the agents is the learning factor, which assumes a double meaning: from one side it represents the speed of learning (i.e. How fast a player change its strategy), from the other it is a sort of risk propensity, that is how much players are encouraged to risk after a winning. Finally we studied how agents adapt their strategy in order to improve their own fitness/gain, by considering the amount of chips as the fitness criterion.
The paradigm of cognitive sciences defines cognitive processes as entities intrinsically dynamical opposed to cognitive functions. While cognitive functions appear as essentially static pattern of information analysis, developed only in certain critical period of human ontogenetical development, cognitive processes are dynamically adaptive and tend to evolve along all human life.
Consequently in order to capture their fundamental features it is necessary to represents toroughly those variables which couple agents and environment. Thus the amount of chips was used both as an indicator of agents' fitness, and a parameter for the evolution of strategies.

First model we proposed is the most general one, and allows a qualitative investigation considering four players and a game structure very closed to the original poker.
The principal insight of first model's numerical simulations deals with the emergence of a coherent and rational strategy to play poker, starting from the very simple scaffholding presented above. All agents learn to play poker and moreover they seem to develop quickly efficient strategies which comprehend rational features as so as bluffing.

The role of learning factor apparently affects deeply agents' fitness. Agents characterized by the same values of learning factor have on average a gain equal to zero, with fluctuations around it vanishing as the number of matches increases.

Instead, when an agent has a higher learning factor value than opponents, its fitness appears on average to be greater than the others (i.e. A greater amount of chips), with fitness which increases with learning factor. Of course the effect is inverted in the opposite case.

The raises distribution shows an exponential behaviour for each values of learning factor ($\mu$), despite this an interesting feature is detectable when different values of learning factor are considered. Maximum raise shows a non trivial behaviour as a function of the LF: it is minimum for small values of LF, then it finds its maximum for an intermediate value of LF and then decreases again for higher values of LF.

Finally an interesting interpretation of results seems to underline the emergence of bluff. Indeed analizing strategy matrices it can be notice some 'irregularities' in raise and call's matrices for all agents, both in the homogenous and heterogeneous case. If we assume as a rational strategy to bet proportionaly to both hand value and pot size, irregularities are represented by deviations from this  behaviour.   

Second model represents the most simple one, but allows to capture the main behaviours of the first model, demonstrating how fundamental is bluffing as rational strategy.

First of all this model confirms general behaviour of the first one. Also in this case the role of LF is preminent. In a straightforward way it's here detectable a direct connection between LF and bluffing tendency. In fact even though both the agents tend to develop bluff, the agent with the greatest LF bluffs more than the other. Finally this last evidence suggests that bluffing is an evolutionary rational strategy.


\

\end{document}